\documentclass[%
 aip,
 amsmath,amssymb,
 reprint,%
]{revtex4-1}

\usepackage{graphicx}
\usepackage{dcolumn}
\usepackage{bm}
\usepackage{color}
\usepackage[utf8]{inputenc}
\usepackage[T1]{fontenc}
\usepackage{mathptmx}

\usepackage{epstopdf}
\usepackage{natbib}
\usepackage{textcomp}

\begin{document}

\preprint{AIP/123-QED}

\title{A versatile and compact high-intensity electron beam for multi-kGy irradiation in nano or micro-electronic devices}

\author{F.Gobet}
 \email{gobet@cenbg.in2p3.fr}
\affiliation{Universit\'e Bordeaux, CNRS-IN2P3, CENBG, F-33175 Gradignan, France}
\author{J.Gardelle}%
\affiliation{CEA, CESTA, F-33116 Le Barp, France}
\author{M.Versteegen}%
\affiliation{Universit\'e Bordeaux, CNRS-IN2P3, CENBG, F-33175 Gradignan, France}
\author{L.Courtois}%
\affiliation{CEA, CESTA, F-33116 Le Barp, France}
\author{S.Leblanc}%
\affiliation{Universit\'e Bordeaux, CNRS-IN2P3, CENBG, F-33175 Gradignan, France}
\author{V.Meot}%
\affiliation{CEA, DIF, F-91297 Arpajon, France}

\date{\today}

\begin{abstract}
A compact low-energy and high-intensity electron source for material aging applications is presented. A laser-induced plasma moves inside a 30 kV diode and produces a 5 MW electron beam at the anode location. The corresponding dose that can be deposited into silicon or gallium samples is estimated to be 25 kGy per laser shot. The dose profile strongly depends on the cathode voltage and can be adjusted from 100 nm to 1 $\mu$m. With this versatile source, a path is opened to study micro or nano-electronic components under high irradiation, without the standard radioprotection issues.
\end{abstract}

\maketitle

The development of resistant materials in harsh radiation environments is of prime importance for the reliability of systems in aerospace, nuclear and military industries\cite{cressler2017extreme}. Hardening studies of micro and nanoscale electronic circuits are currently performed under irradiation doses ranging from 1 Gy to a few tens of kGy\cite{yan2015impact,xu2013nano,fleetwood2013total,fang2014total,cellere2004ionizing,yan2014impact,xi2018impact,tala2015experimental,lee2017low,gaillardin2015high}. Strong alterations of performances are observed for doses larger than 1 kGy. For instance, if the stability of nanometer size memory capacitors  is strongly degraded by irradiation, "writing" operations in memories are prevented\cite{yan2015impact}. The modification of carrier mobility and the reduction of diffusion length have been observed in sub-micrometer scale AlGaN/GaN transistors\cite{lee2017low} leading to severe issues for applications in radiation environments.

Studies of electronic components are generally performed with high energy accelerators or standard radioactive sources such as $^{60}$Co. The energy of the electrons and $\gamma$-rays delivered by these sources is in most cases larger than 1 MeV, thereby reducing the probability of particle interactions into micrometer-size samples. Consequently, strong activities are necessary to reach a few hundred of Gy per hour\cite{yan2015impact}. Then, in order to reach kGy doses in micro or nanoscale devices, very long irradiation times are required, under severe radioprotection conditions. 

With 10 keV electrons, ranges in silicon or gallium samples are of the order of one micrometer\cite{estar}. A versatile and high intensity source emitting electrons in this energy range is therefore promising to study aging of nano or micro-electronic components under very high irradiation doses. In addition, the opportunity of delivering electrons upon request facilitates radiation protection procedures. We recently investigated a new source of electrons at a few keV \cite{raymond2017energy,verst}. In this letter, we detail an important upgrade of the system allowing acceleration of 30 keV electrons, which is now relevant for small scale sample irradiation. We show that doses up to 25 kGy can be obtained with this source operating in single shot.

\begin{figure}[h]
\includegraphics[height=5cm,trim=0 0 0 0, clip=true]{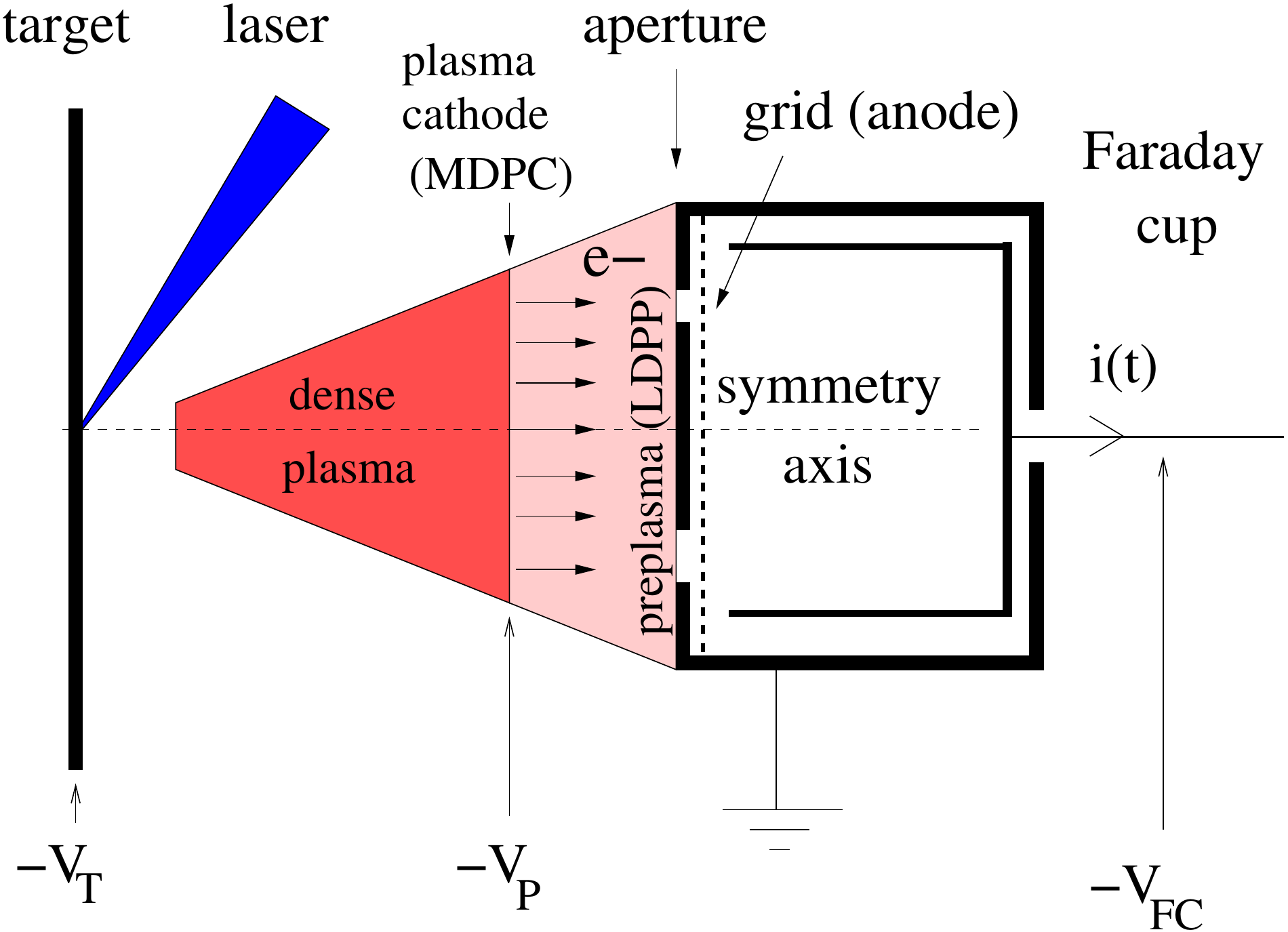}
\caption{Schematic of the compact high-intensity electron source. The plasma positions are indicated some time after the laser shot.}
\end{figure}

The main characteristics of the source were already carefully described\cite{raymond2017energy,verst,comet2016absolute}. They are displayed in Fig.1. A 10 ns, 10$^{13}$ W/cm$^{2}$ Nd:YAG laser pulse is focused on an aluminum target. It produces a plasma in which about 2x10$^{15}$ electrons can be released. This plasma presents two components: a dense aluminum plasma, with density reaching 10$^{20}$-10$^{22}$ part.m$^{-3}$, preceded by a Low Density anisotropic Pre-Plasma (LDPP) with about 10$^{16}$-10$^{17}$ part.m$^{-3}$ expanding in front of the dense plasma. This LDPP contains electrons and light positive ion contaminants of the aluminum target (carbon, nitrogen, hydrogen, and oxygen)\cite{gitomer1986fast}. This two-component plasma expands between the target, biased at a negative voltage -$V_T$, and a grounded 37\% transparent grid (anode) located 50 mm downstream from the target. This grid is inserted at the entrance of a Faraday Cup (FC) which is used to collect the charges and to monitor the beam current. Conducting plates with annular apertures of variable radii are placed in front of the grid. They allow us to measure the current and the charge distributions at the anode location for several distances away from the symmetry axis, up to 28 mm.

Previous studies at $V_T$ = 3 kV and 5 kV showed that during the first 130 ns after the laser shot, electrons are extracted from the front end boundary of the dense plasma component. That interface between the two plasma components acts as a Moving Dense Plasma Cathode\cite{raymond2017energy} (MDPC), biased at a voltage $V_P$ which depends on $V_T$. The mean energy of the extracted electron beam decreases with time as the MDPC expands from the target to the anode. The LDPP in the gap between the MDPC and the anode strongly increases the current of the electron beam\cite{verst}. Actually, the LDPP positive ions change the potential in the gap. A negative electric field appears in front of the MDPP whereas a positive electric field settles in the anode vicinity. The first produces additional acceleration for electrons resulting in greater electron extraction from the MDPC boundary and consequently much higher beam current than would be expected in vacuum. Moreover the anisotropy of the LDPP  is responsible for the inhomogeneous electron beam spatial charge distribution at the anode position\cite{verst} .

When the MDPC fully fills and short-circuits the gap between the target and the anode, a second low energy electron beam of a few hundred eV flows through the anode\cite{raymond2017energy}. Once the MDPC has reached and has short-circuited the anode, a capacitor used to stabilize the target voltage $V_T$ discharges. This capacitor supports the electron current flowing in the cathode-anode gap. These electrons are no longer accelerated by the extracting electric field (short-circuit) and their energy is lower than 1 keV. This low energy flow will not be considered in the rest of this paper.

\begin{figure*}[t]

    \centering
    \begin{minipage}[h]{0.3\linewidth}
     \centering
     \includegraphics[height=6.5cm,trim=0 0 150 80,clip=true]{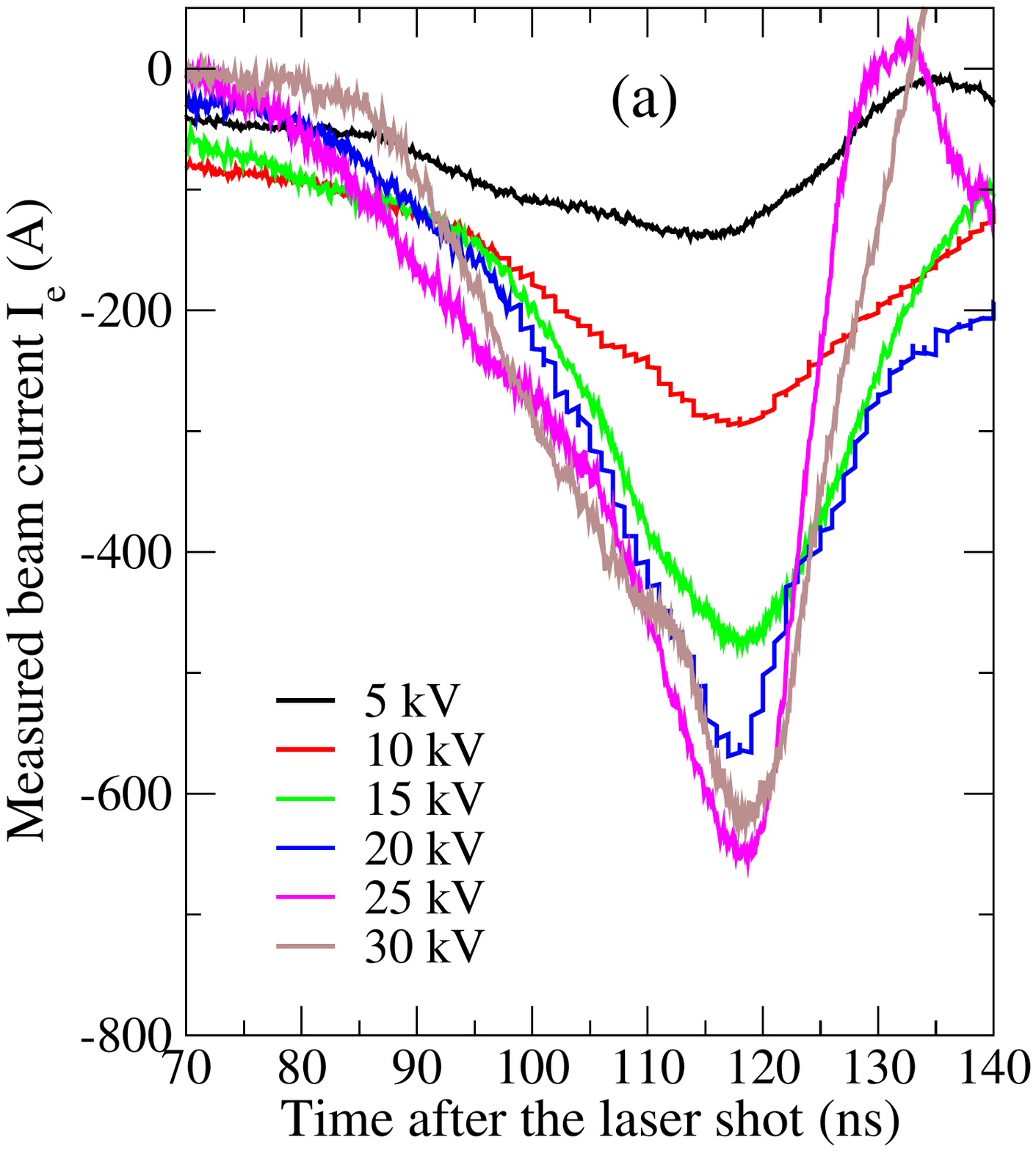}
    \end{minipage}
    \begin{minipage}[h]{0.3\linewidth}
    \centering
        \includegraphics[height=6.5cm,trim=0 0 150 80,clip=true]{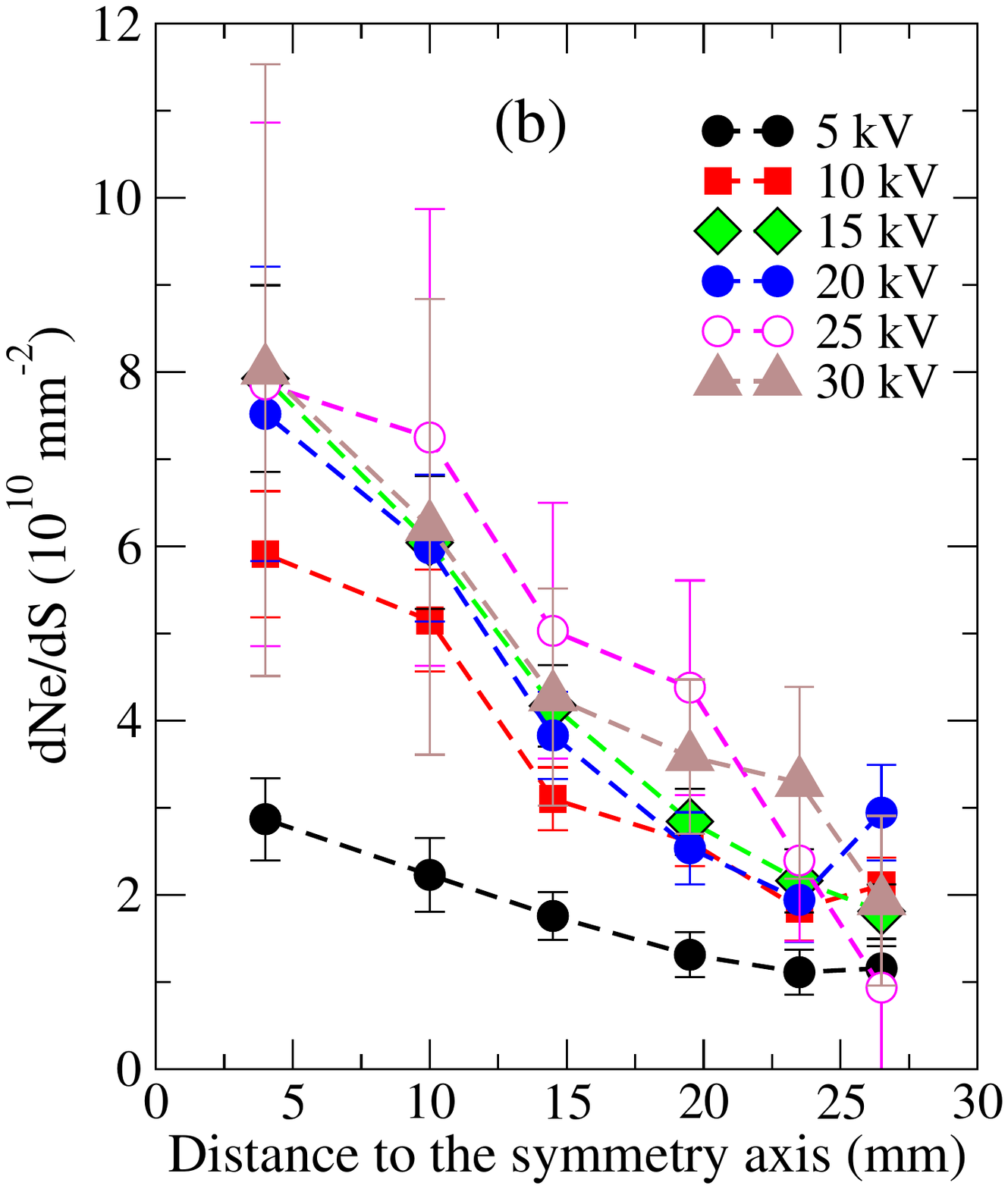}
    \end{minipage}
    \begin{minipage}[h]{0.3\linewidth}
    \centering
        \includegraphics[height=6.5cm,trim=0 0 150 80,clip=true]{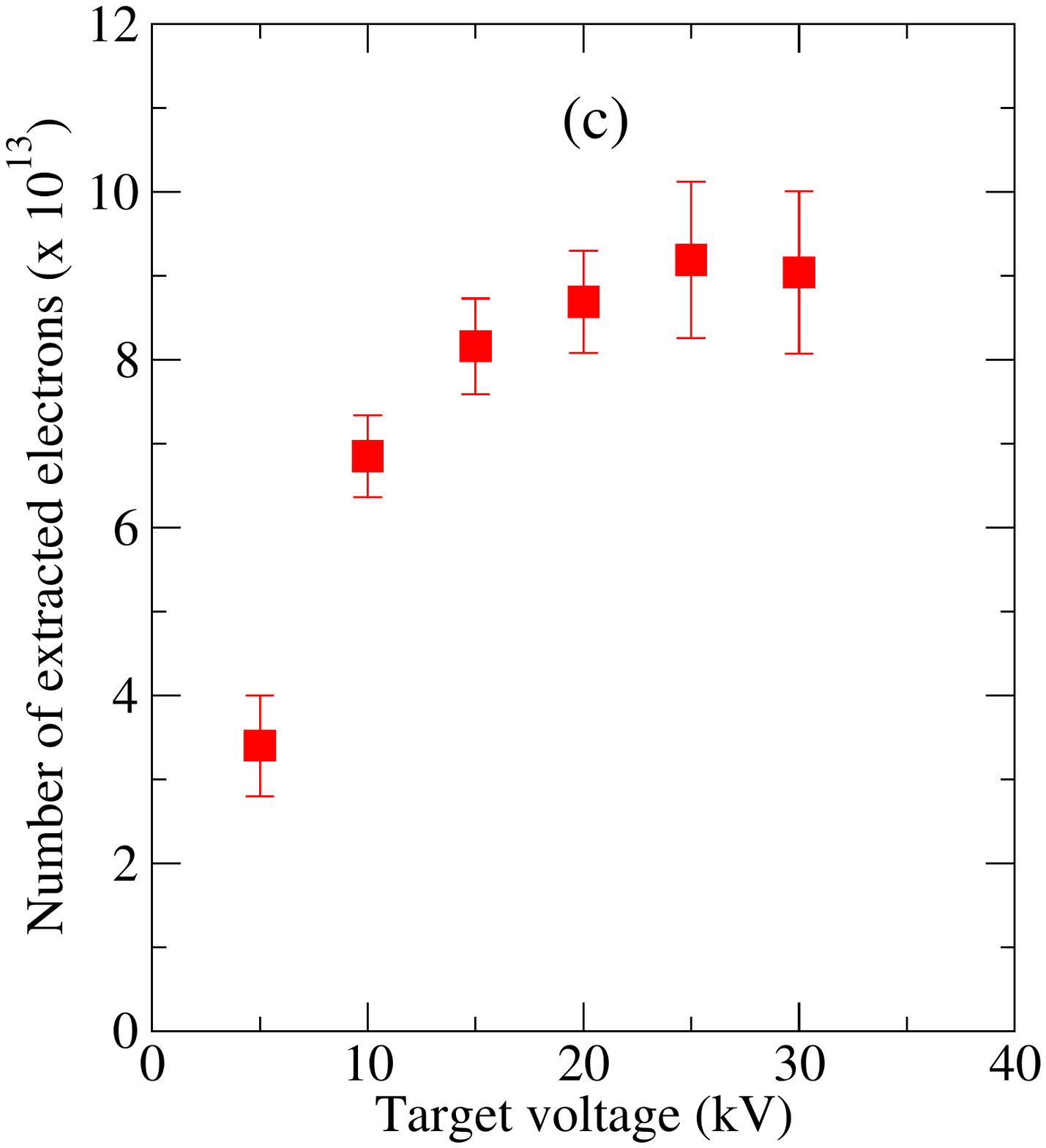}
    \end{minipage}
\caption{a) Measured electron beam current $I_e$ as a function of time after the laser shot. b) Measured charge surface distributions over the first 130 ns after the laser shot. c) Number of extracted electrons reaching the 28 mm radius anode in the first 130 ns after the laser shot, as a function of target voltage. }
\end{figure*}

Figure 2 (a) shows the total beam current $I_e$(t) measured at the anode location, for 6 values of $V_T$.  These curves were obtained by summing the FC current i(t) (see Fig.1) obtained with six different annular apertures covering all radii between 0 and 28 mm. Regardless of $V_T$, the maximum beam current is reached approximately 115-120 ns after the laser shot and, above 20 kV, up to 600 A are collected. If the diode were under vacuum, such a current, much greater than the limiting current of a few amperes, would not propagate\cite{humphries2013charged}. The charge density per unit area of the electrons extracted from the thin MDPC boundary is plotted in Fig.2 (b) as a function of distance to the symmetry axis (see Fig.1) and for the 6 target voltages of Fig.2(a). These distributions are obtained by integrating the corresponding current signals from 0 to 130 ns, this for each annular aperture. The measurements were repeated 10 times and the corresponding statistical uncertainties are represented by the vertical bars. These distributions are strongly inhomogeneous. The collected charge decreases with distance from the symmetry axis and electrons are preferentially extracted in the vicinity of the latter. These results are in agreement with those previously reported at $V_T$ = 3 and 5 kV\cite{verst}. As a first approximation, MDPC expansion and LDPP anisotropy do not depend on the bias voltage value $V_T$.

The total number of electrons collected during the first 130 ns as a function of target voltage is shown in Fig.2 (c). The number of extracted electrons increases with $V_T$ and saturates at 9.10$^{13}$ for 20 kV. This number corresponds to approximately 4.5\% of the 2x10$^{15}$ free electrons present in the plasma. For each value of V$_T$, the beam current is regulated by the excess of positive charge density of the LDPP that fills the gap between the MDPC and the anode\cite{verst}. 

To understand the origin of the electron beam saturation and its dependence with the LDPP properties, simulations are performed by using the 2D axisymmetric particle in cell code XooPIC\cite{verboncoeur1995object}. The geometry is a simple cylinder for which one side corresponds to the MDPC position at time t =110 ns and the other side is the anode grid. The latter is located 8 mm downstream from the MDPC at this particular time. As reported in ref 13, the MDPC voltage $V_P$ is deduced from the target voltage $V_T$. As a first approximation, the ratio between these two voltages is constant and takes the value of 0.4 at t  = 110 ns.

A first series of simulations was performed to estimate the mean positive charge density of the LDPP capable of producing the electron experimental current $I_e$ measured at time t = 110 ns. In this simulation, the gap between the MDPC and the anode is filled with an anisotropic LDPP (protons) whose angular distribution was measured in a previous experiment\cite{verst}. An electron beam is “emitted” with current $I$ from the MDPC which is biased at the potential  $V_P$. Then, the electrons propagate in the LDPP towards the anode and the current $I_A$ that reaches the latter is estimated. The current $I_A$ increases with $I$ up to a maximum value $I_{VC}$ for which a virtual cathode is formed in front of the MDPC, preventing further electron propagation\cite{child1911discharge,langmuir1923effect}. By varying the LDPP positive charge density we have calculated the proton density that produces the experimental current (i.e. $I_{VC}$ = $I_e$), this for each value of V$_P$ (which depends on V$_T$). This density is given by the black circles in Fig. 3, as a function of $V_T$. For a grounded target, the charge density is zero and the  LDPP  is neutral. When the target is polarized, the LDPP positive  charge density necessary to restitute the measured current $I_e $ increases and saturates at 7.2 nC/cm$^{3}$, for target voltages greater than 15 kV. This value is close to the 9 nC/cm$^{3}$ density of positive charges that was measured in the previous experiment\cite{verst} at t = 110 ns. We removed the electrons entering the FC by applying a voltage $V_{FC}$ inside the latter (see Fig.1), such that only positive charges were collected. These simulations suggest that the actual LDPP is gradually emptied of its electrons when the target voltage is increased. Above a voltage of 15 - 20 kV, only positive ions remain in the LDPP, leading to a saturation of the extracted electron current. 

To check this hypothesis, we have performed a second series of simulations. The gap between the MDPC  and the anode is now filled with a homogeneous neutral LDPP including protons and electrons, both with a fixed charge density of 9 nC/cm$^{3}$. When the MDPC is biased at $V_{P}$ (t=110 ns), LDPP electrons move towards the anode. After approximately one nanosecond of calculation, a steady state is reached and the LDPP exhibits an excess of positive charges as shown by the red circles of Fig.3. Electrons are completely removed at 35
 kV
, thus confirming the hypothesis suggested previously.

\begin{figure}[h]
\includegraphics[height=6cm,trim=0 40 400 200, clip=true]{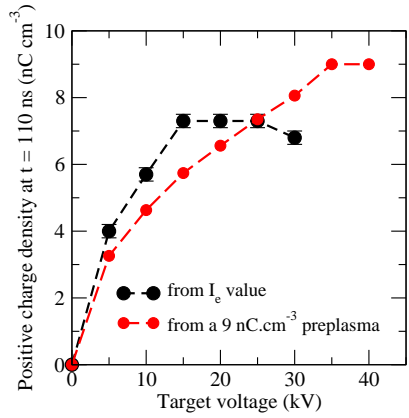}
\caption{Dependence of the excess of positive 
charge density in the LDPP with the target voltage, $t=110$ ns after the laser shot. Black circles: calculations constrained by the measured beam current. Red circles: calculations of the steady state for a neutral LDPP (electrons+protons) having a charge density of 9 nC.cm$^{-3}$ for both species.}
\end{figure}

The experimental energy distribution of the electrons, as obtained from the beam current $I_e(t)$ and $V_{P}(t)$, is given by:
\begin{eqnarray}
 \frac{dN}{dE} (E) = \sum_{t=0}^{130\ ns} \frac{I_e(t)\Delta t}{-e} \delta(-e V_{P}(t)-E\pm\Delta E/2)
\end{eqnarray}
where $\delta$ is a function defined as $\delta (-eV_{P}(t)-E\pm \Delta E/2)=1$ if $E-\Delta E/2\leq -eV_{P}(t) \leq E+\Delta E/2$ and 0 otherwise. We have considered a sampling interval $\Delta t=100$ ps and an energy bin $\Delta E=500$ eV. The energy distributions are displayed in Fig. 4 for the six target voltages of operation. They are continuous and indicate a maximum energy greater than $eV_T$. During the first times of extraction, electrons are accelerated toward the anode by the electric field induced by $V_T$ but they are also repelled by their followers, allowing them to gain additional kinetic energy.
Depending on the target voltage, the average energy of the extracted electron beam varies from about 1 to 10 keV. In this way, the extracted electrons can deposit their energy into thin (<$\mu$m) silicon or gallium samples such as those used in nano or microelectronic structures.

\begin{figure}[h]
\includegraphics[height=8.cm,trim=0 60 200 20, clip=true]{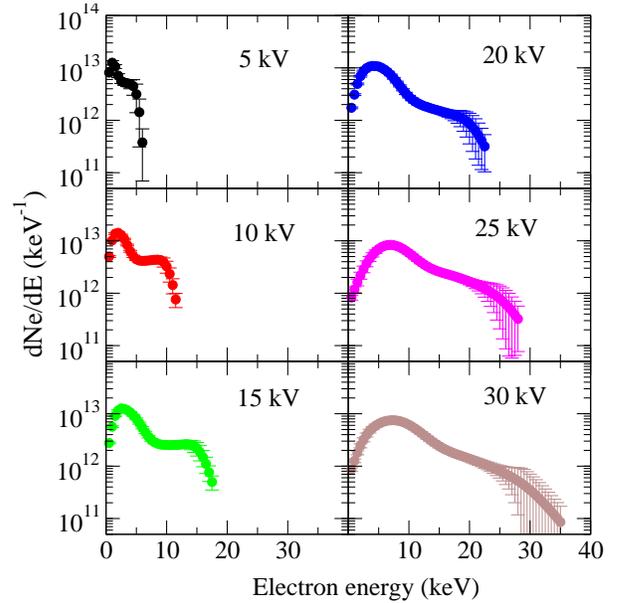}
\caption{Experiment: energy distributions of the electrons measured between 0 and 130 ns when they have reached the 24.6 cm$^2$ anode area, at different target voltages.}
\end{figure}

The beam peak power $I_eV_P$ is displayed in Fig.5(a) for three values of $V_T$. If we average these curves between t = 80 and 130 ns, we observe the approximately quadratic behavior with V$_T$ shown in Fig.5 (b).  We can estimate the kinetic energy transported by the electron beam over the 24 cm$^2$ anode area. Assuming 5 MW of peak-power, as obtained at 30 kV, and a deposition which lasts 25 ns, as shown in Fig.5 (a), we obtain an energy of 100 mJ.  If this energy were deposited in a 1 mm slice of silicon ($\rho$=2300 $kg/m^3$), the dose would be 18 kGy, a value much higher than yields obtained with conventional radioactive sources. In addition this dose  is due to a single laser-plasma shot.

\begin{figure}[h]
    \centering
    \begin{minipage}[h]{0.9\linewidth}
     \centering
     \includegraphics[height=4.2cm,trim=40 44 300 216,clip=true]{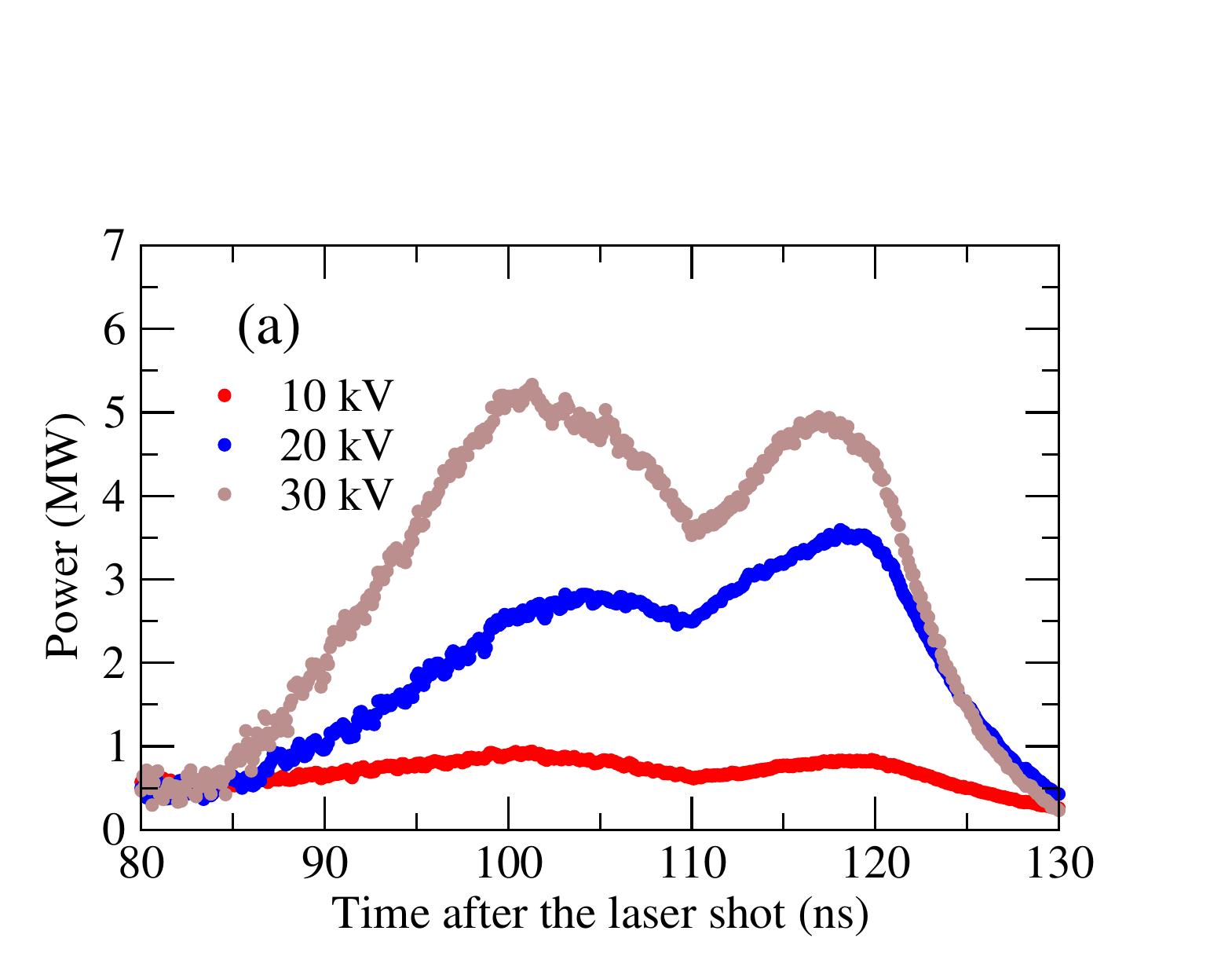}\vspace{1.cm}\par
      \includegraphics[height=5.cm,trim=0 40 200 80, clip=true]{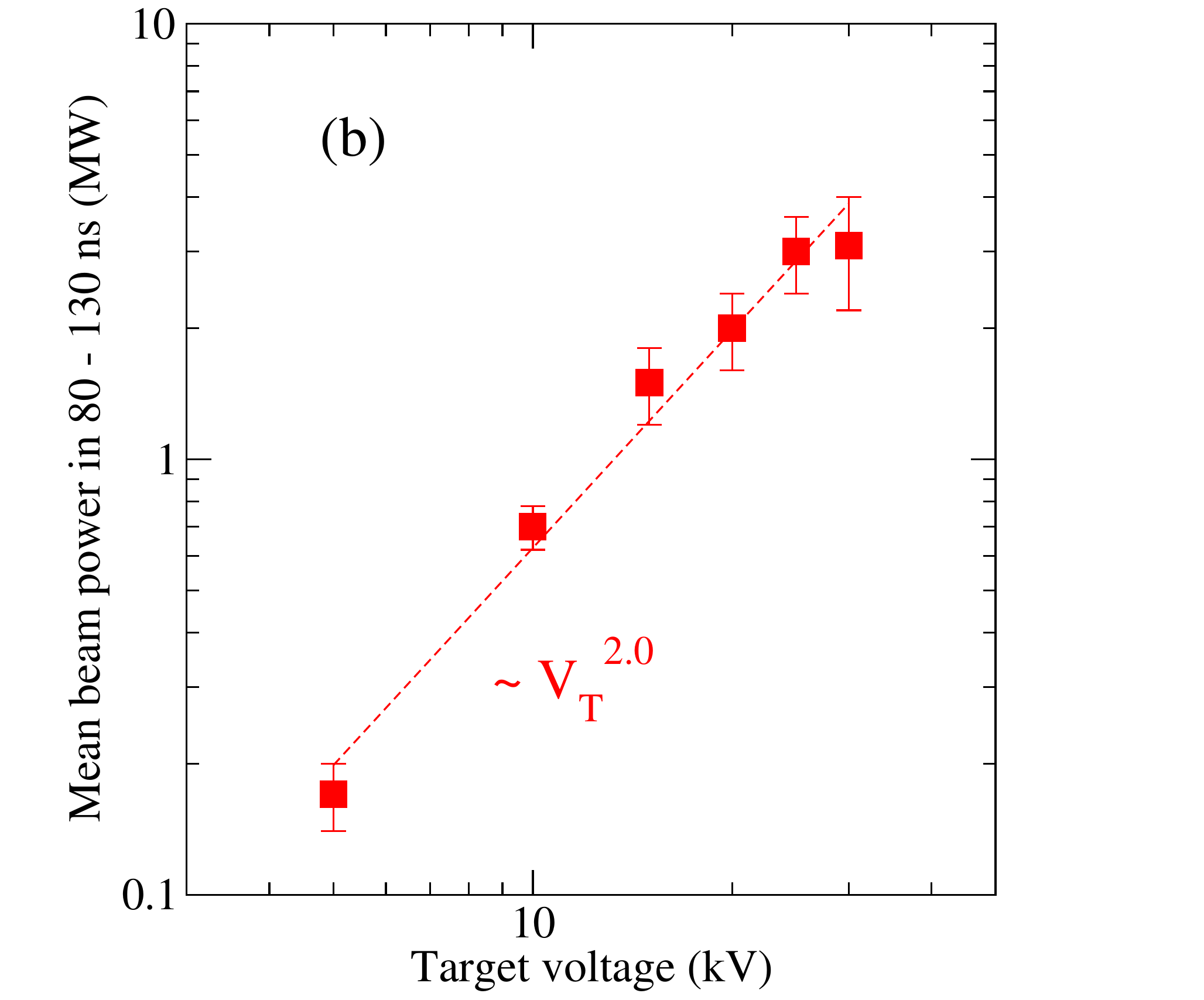}
    \end{minipage}

\caption{a) Electron beam power as a function of time for three target voltages V$_T$   b) Electron beam maximum power as a function of the target voltage.}
\end{figure}

Let us now specify a simple way to use this source as an irradiation facility. First of all, we remove the Faraday Cup and replace the grid by a conducting anode. We select the central and homogenous part of the electron beam by inserting a 1 mm radius aperture in the anode center. Then, we place a 100 nm-thick silicon nitride window to stop the secondary low energy electrons coming from the capacitor discharge but also the plasma ions. Finally, we install the sample under study just after this filter. The beam deposition in the sample can be estimated by using the Monte Carlo code Geant4\cite{agostinelli2003geant4} associated with the Livermore Physics List\cite{allison2016recent}. This allowed us to compute particle interactions (electrons, photons) with energies as low as 100 eV. Multiple scattering, Bremsstrahlung and continuous energy loss are taken into account for electrons. The physical processes for photons are the Compton and Photoelectric effects and Rayleigh scattering. In addition, secondary electron emission and fluorescence are also included. 

The simulation is initialized with an electron beam having the characteristics reported above. We have computed the depth dose profile in key materials, such as silicon and gallium, for different target voltages $V_T$. We have considered 10 mm thick samples of radius 1 mm. For each target voltage, 10$^9$ electrons are sent perpendicular to the sample. They have the measured energy distributions reported in Fig.4. The energy deposited by the primary and the secondary particles in the sample is recorded as a function of its depth, with a resolution of 1 nm. The dose profile is obtained by dividing the total energy deposited in each 1 nm-thick disk by its mass. Then, it is rescaled to the experimental number of electrons passing through the $\pi$ mm$^2$ aperture area, accordingly to the results presented in Fig. 2(b).

\begin{figure}

    \centering
    \begin{minipage}[h]{0.49\linewidth}
     \centering
     \includegraphics[height=6.2cm,trim=20 0 20 0,clip=true]{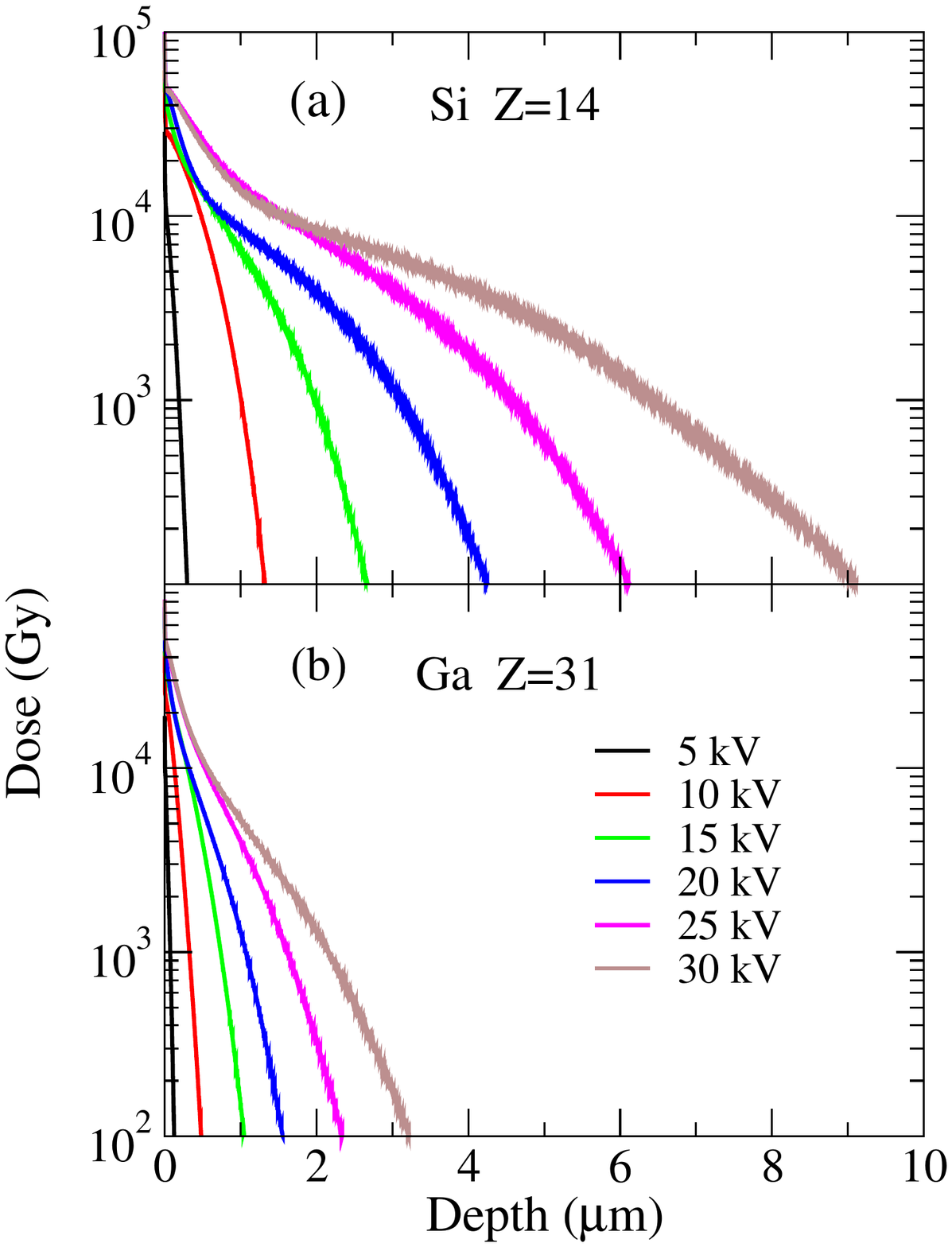}

    \end{minipage}\hspace{0.cm}
    \begin{minipage}[h]{0.5\linewidth}
    \centering
        \includegraphics[height=6.2cm,trim=0 0 150 0,clip=true]{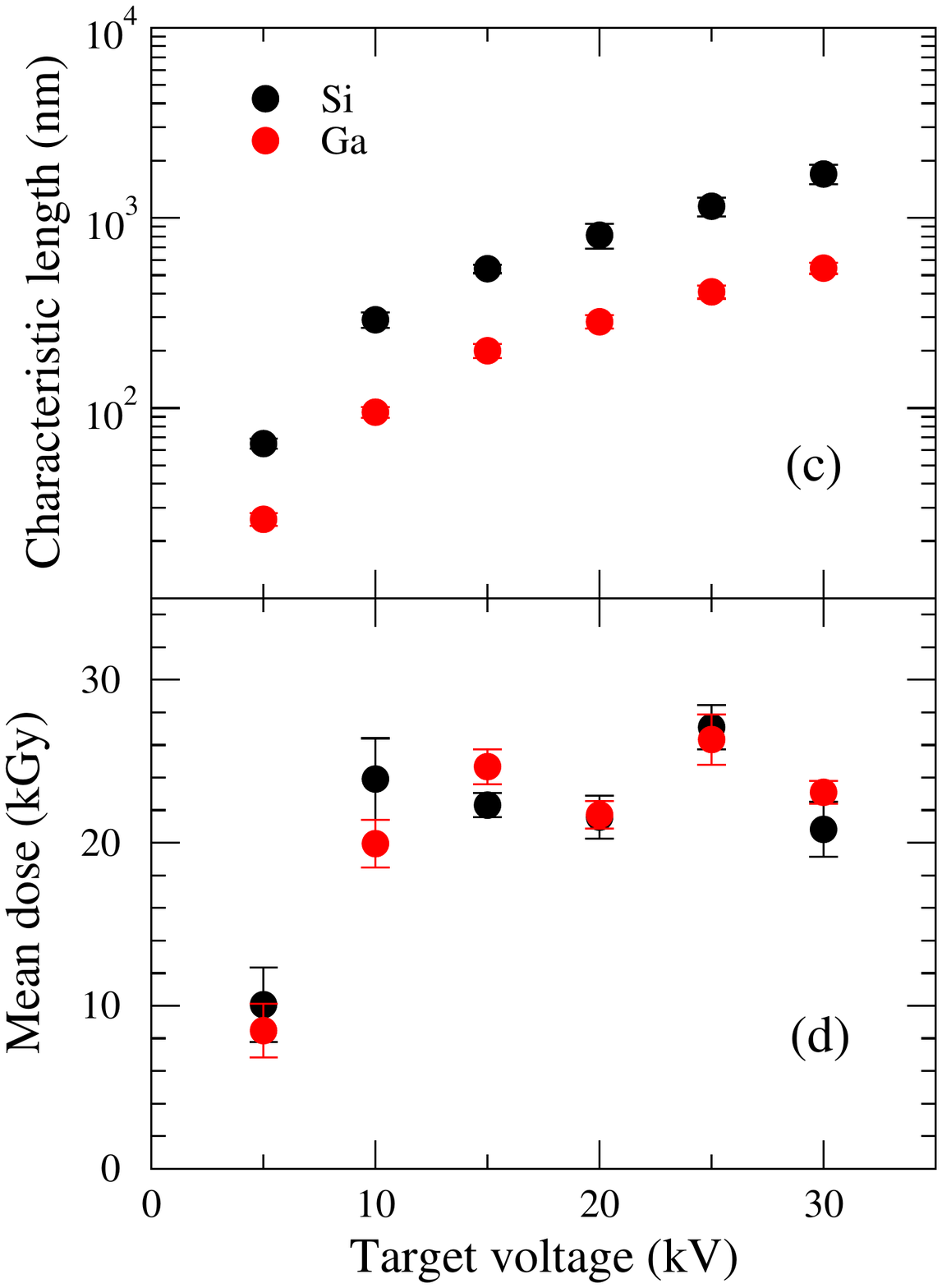}
    \end{minipage}
\caption{Geant4 calculations of dose profiles in a) silicon, b) gallium samples for different target voltages, c) characteristic length l$_p$, d) mean dose averaged over one $l_p$.}
\end{figure} 

The results of these simulations for silicon and gallium samples are shown in Fig. 6(a) and 6(b) respectively. The curves represent the dose profiles $D(x)$ for our six usual values of $V_T$. A maximum dose of 40 kGy is obtained, with a dose depth of order 1 $\mu$m. As shown in these figures, the key feature of this new electron source is the capability of modulating the depth of energy deposition by varying the target voltage. 
We have defined a characteristic length as 
\begin{eqnarray}
l_p=\frac{\int xD(x)\,dx}{\int D(x)\,dx}.
\end{eqnarray}

This length $l_p$, and the corresponding average dose deposited inside, are reported in Fig. 6(c) and 6(d) as a function of $V_T$. We observe that the average dose, of order 25 kGy, is basically independent of the target voltage and material for electron kinetic energy greater than 10 keV. However, the dose depth strongly depends on the target voltage and increases from a few tens of nm at 5 kV up to one micrometer at 30 kV. The length $l_p$ is slightly lower for gallium than for silicon as its atomic number is higher. Additional calculations on tantalum (Z=73) have given a length $l_p$ of the order of 200 nm at 30 kV. Therefore, for a nano or microelectronic device aging application, the target voltage can be easily chosen according to the material thickness and composition.

In summary, we have analyzed the characteristics of the electrons extracted from a laser-induced plasma diode up to 30 kV. Saturation of the collected number of electrons is observed when the applied target voltage reaches 20 kV, as a result of the full depletion of the LDPP produced during laser-target interaction. This new versatile and compact electron source is therefore relevant for micro or nanotechnology aging studies under high irradiation dose. Typically, with a laser repetition rate of 10 Hz, a dose of the order of 1 MGy could be available in less than 5 s. Radioprotection issues, inherent to high radioactivity in standard sources, can be addressed. This new electron source is also useful for irradiation of biological samples. For example, fragmentation yields of hydrated or dried plasmids under high dose exposition\cite{souici2017single,hahn2017measurements} could be studied.

\bibliographystyle{apsrev4-1}

\bibliography{biblio}

\begin{thebibliography}{24}%
\makeatletter
\providecommand \@ifxundefined [1]{%
 \@ifx{#1\undefined}
}%
\providecommand \@ifnum [1]{%
 \ifnum #1\expandafter \@firstoftwo
 \else \expandafter \@secondoftwo
 \fi
}%
\providecommand \@ifx [1]{%
 \ifx #1\expandafter \@firstoftwo
 \else \expandafter \@secondoftwo
 \fi
}%
\providecommand \natexlab [1]{#1}%
\providecommand \enquote  [1]{``#1''}%
\providecommand \bibnamefont  [1]{#1}%
\providecommand \bibfnamefont [1]{#1}%
\providecommand \citenamefont [1]{#1}%
\providecommand \href@noop [0]{\@secondoftwo}%
\providecommand \href [0]{\begingroup \@sanitize@url \@href}%
\providecommand \@href[1]{\@@startlink{#1}\@@href}%
\providecommand \@@href[1]{\endgroup#1\@@endlink}%
\providecommand \@sanitize@url [0]{\catcode `\\12\catcode `\$12\catcode
  `\&12\catcode `\#12\catcode `\^12\catcode `\_12\catcode `\%12\relax}%
\providecommand \@@startlink[1]{}%
\providecommand \@@endlink[0]{}%
\providecommand \url  [0]{\begingroup\@sanitize@url \@url }%
\providecommand \@url [1]{\endgroup\@href {#1}{\urlprefix }}%
\providecommand \urlprefix  [0]{URL }%
\providecommand \Eprint [0]{\href }%
\providecommand \doibase [0]{http://dx.doi.org/}%
\providecommand \selectlanguage [0]{\@gobble}%
\providecommand \bibinfo  [0]{\@secondoftwo}%
\providecommand \bibfield  [0]{\@secondoftwo}%
\providecommand \translation [1]{[#1]}%
\providecommand \BibitemOpen [0]{}%
\providecommand \bibitemStop [0]{}%
\providecommand \bibitemNoStop [0]{.\EOS\space}%
\providecommand \EOS [0]{\spacefactor3000\relax}%
\providecommand \BibitemShut  [1]{\csname bibitem#1\endcsname}%
\let\auto@bib@innerbib\@empty
\bibitem [{\citenamefont {Cressler}\ and\ \citenamefont
  {Mantooth}(2017)}]{cressler2017extreme}%
  \BibitemOpen
  \bibfield  {author} {\bibinfo {author} {\bibfnamefont {J.~D.}\ \bibnamefont
  {Cressler}}\ and\ \bibinfo {author} {\bibfnamefont {H.~A.}\ \bibnamefont
  {Mantooth}},\ }\href@noop {} {\emph {\bibinfo {title} {Extreme environment
  electronics}}}\ (\bibinfo  {publisher} {CRC Press},\ \bibinfo {year}
  {2017})\BibitemShut {NoStop}%
\bibitem [{\citenamefont {Yan}\ \emph {et~al.}(2015)\citenamefont {Yan},
  \citenamefont {Zhao}, \citenamefont {Guo}, \citenamefont {Xiong},
  \citenamefont {Tang}, \citenamefont {Li}, \citenamefont {Xiao}, \citenamefont
  {Zhang}, \citenamefont {Ding}, \citenamefont {Chen} \emph
  {et~al.}}]{yan2015impact}%
  \BibitemOpen
  \bibfield  {author} {\bibinfo {author} {\bibfnamefont {S.}~\bibnamefont
  {Yan}}, \bibinfo {author} {\bibfnamefont {W.}~\bibnamefont {Zhao}}, \bibinfo
  {author} {\bibfnamefont {H.}~\bibnamefont {Guo}}, \bibinfo {author}
  {\bibfnamefont {Y.}~\bibnamefont {Xiong}}, \bibinfo {author} {\bibfnamefont
  {M.}~\bibnamefont {Tang}}, \bibinfo {author} {\bibfnamefont {Z.}~\bibnamefont
  {Li}}, \bibinfo {author} {\bibfnamefont {Y.}~\bibnamefont {Xiao}}, \bibinfo
  {author} {\bibfnamefont {W.}~\bibnamefont {Zhang}}, \bibinfo {author}
  {\bibfnamefont {H.}~\bibnamefont {Ding}}, \bibinfo {author} {\bibfnamefont
  {J.}~\bibnamefont {Chen}},  \emph {et~al.},\ }\href@noop {} {\bibfield
  {journal} {\bibinfo  {journal} {Applied Physics Letters}\ }\textbf {\bibinfo
  {volume} {106}},\ \bibinfo {pages} {012901} (\bibinfo {year}
  {2015})}\BibitemShut {NoStop}%
\bibitem [{\citenamefont {Xu}\ \emph {et~al.}(2013)\citenamefont {Xu},
  \citenamefont {Chen}, \citenamefont {Zhou}, \citenamefont {Li}, \citenamefont
  {Li}, \citenamefont {Niu}, \citenamefont {Shan}, \citenamefont {Guo},
  \citenamefont {Wang},\ and\ \citenamefont {Qian}}]{xu2013nano}%
  \BibitemOpen
  \bibfield  {author} {\bibinfo {author} {\bibfnamefont {Z.}~\bibnamefont
  {Xu}}, \bibinfo {author} {\bibfnamefont {L.}~\bibnamefont {Chen}}, \bibinfo
  {author} {\bibfnamefont {B.}~\bibnamefont {Zhou}}, \bibinfo {author}
  {\bibfnamefont {Y.}~\bibnamefont {Li}}, \bibinfo {author} {\bibfnamefont
  {B.}~\bibnamefont {Li}}, \bibinfo {author} {\bibfnamefont {J.}~\bibnamefont
  {Niu}}, \bibinfo {author} {\bibfnamefont {M.}~\bibnamefont {Shan}}, \bibinfo
  {author} {\bibfnamefont {Q.}~\bibnamefont {Guo}}, \bibinfo {author}
  {\bibfnamefont {Z.}~\bibnamefont {Wang}}, \ and\ \bibinfo {author}
  {\bibfnamefont {X.}~\bibnamefont {Qian}},\ }\href@noop {} {\bibfield
  {journal} {\bibinfo  {journal} {Rsc Advances}\ }\textbf {\bibinfo {volume}
  {3}},\ \bibinfo {pages} {10579} (\bibinfo {year} {2013})}\BibitemShut
  {NoStop}%
\bibitem [{\citenamefont {Fleetwood}(2013)}]{fleetwood2013total}%
  \BibitemOpen
  \bibfield  {author} {\bibinfo {author} {\bibfnamefont {D.~M.}\ \bibnamefont
  {Fleetwood}},\ }\href@noop {} {\bibfield  {journal} {\bibinfo  {journal}
  {IEEE Transactions on Nuclear Science}\ }\textbf {\bibinfo {volume} {60}},\
  \bibinfo {pages} {1706} (\bibinfo {year} {2013})}\BibitemShut {NoStop}%
\bibitem [{\citenamefont {Fang}\ \emph {et~al.}(2014)\citenamefont {Fang},
  \citenamefont {Gonzalez~Velo}, \citenamefont {Chen}, \citenamefont {Holbert},
  \citenamefont {Kozicki}, \citenamefont {Barnaby},\ and\ \citenamefont
  {Yu}}]{fang2014total}%
  \BibitemOpen
  \bibfield  {author} {\bibinfo {author} {\bibfnamefont {R.}~\bibnamefont
  {Fang}}, \bibinfo {author} {\bibfnamefont {Y.}~\bibnamefont {Gonzalez~Velo}},
  \bibinfo {author} {\bibfnamefont {W.}~\bibnamefont {Chen}}, \bibinfo {author}
  {\bibfnamefont {K.~E.}\ \bibnamefont {Holbert}}, \bibinfo {author}
  {\bibfnamefont {M.~N.}\ \bibnamefont {Kozicki}}, \bibinfo {author}
  {\bibfnamefont {H.}~\bibnamefont {Barnaby}}, \ and\ \bibinfo {author}
  {\bibfnamefont {S.}~\bibnamefont {Yu}},\ }\href@noop {} {\bibfield  {journal}
  {\bibinfo  {journal} {Applied Physics Letters}\ }\textbf {\bibinfo {volume}
  {104}},\ \bibinfo {pages} {183507} (\bibinfo {year} {2014})}\BibitemShut
  {NoStop}%
\bibitem [{\citenamefont {Cellere}\ \emph {et~al.}(2004)\citenamefont
  {Cellere}, \citenamefont {Paccagnella}, \citenamefont {Visconti},\ and\
  \citenamefont {Bonanomi}}]{cellere2004ionizing}%
  \BibitemOpen
  \bibfield  {author} {\bibinfo {author} {\bibfnamefont {G.}~\bibnamefont
  {Cellere}}, \bibinfo {author} {\bibfnamefont {A.}~\bibnamefont
  {Paccagnella}}, \bibinfo {author} {\bibfnamefont {A.}~\bibnamefont
  {Visconti}}, \ and\ \bibinfo {author} {\bibfnamefont {M.}~\bibnamefont
  {Bonanomi}},\ }\href@noop {} {\bibfield  {journal} {\bibinfo  {journal}
  {Applied Physics Letters}\ }\textbf {\bibinfo {volume} {85}},\ \bibinfo
  {pages} {485} (\bibinfo {year} {2004})}\BibitemShut {NoStop}%
\bibitem [{\citenamefont {Yan}\ \emph {et~al.}(2014)\citenamefont {Yan},
  \citenamefont {Xiong}, \citenamefont {Tang}, \citenamefont {Li},
  \citenamefont {Xiao}, \citenamefont {Zhang}, \citenamefont {Zhao},
  \citenamefont {Guo}, \citenamefont {Ding}, \citenamefont {Chen} \emph
  {et~al.}}]{yan2014impact}%
  \BibitemOpen
  \bibfield  {author} {\bibinfo {author} {\bibfnamefont {S.}~\bibnamefont
  {Yan}}, \bibinfo {author} {\bibfnamefont {Y.}~\bibnamefont {Xiong}}, \bibinfo
  {author} {\bibfnamefont {M.}~\bibnamefont {Tang}}, \bibinfo {author}
  {\bibfnamefont {Z.}~\bibnamefont {Li}}, \bibinfo {author} {\bibfnamefont
  {Y.}~\bibnamefont {Xiao}}, \bibinfo {author} {\bibfnamefont {W.}~\bibnamefont
  {Zhang}}, \bibinfo {author} {\bibfnamefont {W.}~\bibnamefont {Zhao}},
  \bibinfo {author} {\bibfnamefont {H.}~\bibnamefont {Guo}}, \bibinfo {author}
  {\bibfnamefont {H.}~\bibnamefont {Ding}}, \bibinfo {author} {\bibfnamefont
  {J.}~\bibnamefont {Chen}},  \emph {et~al.},\ }\href@noop {} {\bibfield
  {journal} {\bibinfo  {journal} {Journal of Applied Physics}\ }\textbf
  {\bibinfo {volume} {115}},\ \bibinfo {pages} {204504} (\bibinfo {year}
  {2014})}\BibitemShut {NoStop}%
\bibitem [{\citenamefont {Xi}\ \emph {et~al.}(2018)\citenamefont {Xi},
  \citenamefont {Bi}, \citenamefont {Hu}, \citenamefont {Li}, \citenamefont
  {Liu}, \citenamefont {Xu},\ and\ \citenamefont {Liu}}]{xi2018impact}%
  \BibitemOpen
  \bibfield  {author} {\bibinfo {author} {\bibfnamefont {K.}~\bibnamefont
  {Xi}}, \bibinfo {author} {\bibfnamefont {J.}~\bibnamefont {Bi}}, \bibinfo
  {author} {\bibfnamefont {Y.}~\bibnamefont {Hu}}, \bibinfo {author}
  {\bibfnamefont {B.}~\bibnamefont {Li}}, \bibinfo {author} {\bibfnamefont
  {J.}~\bibnamefont {Liu}}, \bibinfo {author} {\bibfnamefont {Y.}~\bibnamefont
  {Xu}}, \ and\ \bibinfo {author} {\bibfnamefont {M.}~\bibnamefont {Liu}},\
  }\href@noop {} {\bibfield  {journal} {\bibinfo  {journal} {Applied Physics
  Letters}\ }\textbf {\bibinfo {volume} {113}},\ \bibinfo {pages} {164103}
  (\bibinfo {year} {2018})}\BibitemShut {NoStop}%
\bibitem [{\citenamefont {Tala-Ighil}\ \emph {et~al.}(2015)\citenamefont
  {Tala-Ighil}, \citenamefont {Trolet}, \citenamefont {Gualous}, \citenamefont
  {Mary},\ and\ \citenamefont {Lefebvre}}]{tala2015experimental}%
  \BibitemOpen
  \bibfield  {author} {\bibinfo {author} {\bibfnamefont {B.}~\bibnamefont
  {Tala-Ighil}}, \bibinfo {author} {\bibfnamefont {J.-L.}\ \bibnamefont
  {Trolet}}, \bibinfo {author} {\bibfnamefont {H.}~\bibnamefont {Gualous}},
  \bibinfo {author} {\bibfnamefont {P.}~\bibnamefont {Mary}}, \ and\ \bibinfo
  {author} {\bibfnamefont {S.}~\bibnamefont {Lefebvre}},\ }\href@noop {}
  {\bibfield  {journal} {\bibinfo  {journal} {Microelectronics Reliability}\
  }\textbf {\bibinfo {volume} {55}},\ \bibinfo {pages} {1512} (\bibinfo {year}
  {2015})}\BibitemShut {NoStop}%
\bibitem [{\citenamefont {Lee}\ \emph {et~al.}(2017)\citenamefont {Lee},
  \citenamefont {Yadav}, \citenamefont {Antia}, \citenamefont {Zaffino},
  \citenamefont {Flitsiyan}, \citenamefont {Chernyak}, \citenamefont {Salzman},
  \citenamefont {Meyler}, \citenamefont {Ahn}, \citenamefont {Ren} \emph
  {et~al.}}]{lee2017low}%
  \BibitemOpen
  \bibfield  {author} {\bibinfo {author} {\bibfnamefont {J.}~\bibnamefont
  {Lee}}, \bibinfo {author} {\bibfnamefont {A.}~\bibnamefont {Yadav}}, \bibinfo
  {author} {\bibfnamefont {M.}~\bibnamefont {Antia}}, \bibinfo {author}
  {\bibfnamefont {V.}~\bibnamefont {Zaffino}}, \bibinfo {author} {\bibfnamefont
  {E.}~\bibnamefont {Flitsiyan}}, \bibinfo {author} {\bibfnamefont
  {L.}~\bibnamefont {Chernyak}}, \bibinfo {author} {\bibfnamefont
  {J.}~\bibnamefont {Salzman}}, \bibinfo {author} {\bibfnamefont
  {B.}~\bibnamefont {Meyler}}, \bibinfo {author} {\bibfnamefont
  {S.}~\bibnamefont {Ahn}}, \bibinfo {author} {\bibfnamefont {F.}~\bibnamefont
  {Ren}},  \emph {et~al.},\ }\href@noop {} {\bibfield  {journal} {\bibinfo
  {journal} {Radiation Effects and Defects in Solids}\ }\textbf {\bibinfo
  {volume} {172}},\ \bibinfo {pages} {250} (\bibinfo {year}
  {2017})}\BibitemShut {NoStop}%
\bibitem [{\citenamefont {Gaillardin}\ \emph {et~al.}(2015)\citenamefont
  {Gaillardin}, \citenamefont {Martinez}, \citenamefont {Girard}, \citenamefont
  {Goiffon}, \citenamefont {Paillet}, \citenamefont {Leray}, \citenamefont
  {Magnan}, \citenamefont {Ouerdane}, \citenamefont {Boukenter}, \citenamefont
  {Marcandella} \emph {et~al.}}]{gaillardin2015high}%
  \BibitemOpen
  \bibfield  {author} {\bibinfo {author} {\bibfnamefont {M.}~\bibnamefont
  {Gaillardin}}, \bibinfo {author} {\bibfnamefont {M.}~\bibnamefont
  {Martinez}}, \bibinfo {author} {\bibfnamefont {S.}~\bibnamefont {Girard}},
  \bibinfo {author} {\bibfnamefont {V.}~\bibnamefont {Goiffon}}, \bibinfo
  {author} {\bibfnamefont {P.}~\bibnamefont {Paillet}}, \bibinfo {author}
  {\bibfnamefont {J.-L.}\ \bibnamefont {Leray}}, \bibinfo {author}
  {\bibfnamefont {P.}~\bibnamefont {Magnan}}, \bibinfo {author} {\bibfnamefont
  {Y.}~\bibnamefont {Ouerdane}}, \bibinfo {author} {\bibfnamefont
  {A.}~\bibnamefont {Boukenter}}, \bibinfo {author} {\bibfnamefont
  {C.}~\bibnamefont {Marcandella}},  \emph {et~al.},\ }\href@noop {} {\bibfield
   {journal} {\bibinfo  {journal} {IEEE Transactions on Nuclear Science}\
  }\textbf {\bibinfo {volume} {62}},\ \bibinfo {pages} {1226} (\bibinfo {year}
  {2015})}\BibitemShut {NoStop}%
\bibitem [{est()}]{estar}%
  \BibitemOpen
  \href@noop {} {\enquote {\bibinfo {title} {Estar},}\ }\bibinfo {howpublished}
  {\url{http://physics.nist.gov/Star}},\ \bibinfo {note} {january 01,
  1999}\BibitemShut {NoStop}%
\bibitem [{\citenamefont {Raymond}\ \emph {et~al.}(2017)\citenamefont
  {Raymond}, \citenamefont {Versteegen}, \citenamefont {Gobet}, \citenamefont
  {Hannachi}, \citenamefont {Henares},\ and\ \citenamefont
  {Tarisien}}]{raymond2017energy}%
  \BibitemOpen
  \bibfield  {author} {\bibinfo {author} {\bibfnamefont {X.}~\bibnamefont
  {Raymond}}, \bibinfo {author} {\bibfnamefont {M.}~\bibnamefont {Versteegen}},
  \bibinfo {author} {\bibfnamefont {F.}~\bibnamefont {Gobet}}, \bibinfo
  {author} {\bibfnamefont {F.}~\bibnamefont {Hannachi}}, \bibinfo {author}
  {\bibfnamefont {J.~L.}\ \bibnamefont {Henares}}, \ and\ \bibinfo {author}
  {\bibfnamefont {M.}~\bibnamefont {Tarisien}},\ }\href@noop {} {\bibfield
  {journal} {\bibinfo  {journal} {Journal of Applied Physics}\ }\textbf
  {\bibinfo {volume} {122}},\ \bibinfo {pages} {173302} (\bibinfo {year}
  {2017})}\BibitemShut {NoStop}%
\bibitem [{\citenamefont {Versteegen}\ \emph {et~al.}(2019)\citenamefont
  {Versteegen}, \citenamefont {Raymond}, \citenamefont {Gobet},\ and\
  \citenamefont {Henares}}]{verst}%
  \BibitemOpen
  \bibfield  {author} {\bibinfo {author} {\bibfnamefont {M.}~\bibnamefont
  {Versteegen}}, \bibinfo {author} {\bibfnamefont {X.}~\bibnamefont {Raymond}},
  \bibinfo {author} {\bibfnamefont {F.}~\bibnamefont {Gobet}}, \ and\ \bibinfo
  {author} {\bibfnamefont {J.~L.}\ \bibnamefont {Henares}},\ }\href@noop {}
  {\bibfield  {journal} {\bibinfo  {journal} {Review of Scientific
  Instruments}\ }\textbf {\bibinfo {volume} {90}},\ \bibinfo {pages} {053306}
  (\bibinfo {year} {2019})}\BibitemShut {NoStop}%
\bibitem [{\citenamefont {Comet}\ \emph {et~al.}(2016)\citenamefont {Comet},
  \citenamefont {Versteegen}, \citenamefont {Gobet}, \citenamefont
  {Denis-Petit}, \citenamefont {Hannachi}, \citenamefont {Meot},\ and\
  \citenamefont {Tarisien}}]{comet2016absolute}%
  \BibitemOpen
  \bibfield  {author} {\bibinfo {author} {\bibfnamefont {M.}~\bibnamefont
  {Comet}}, \bibinfo {author} {\bibfnamefont {M.}~\bibnamefont {Versteegen}},
  \bibinfo {author} {\bibfnamefont {F.}~\bibnamefont {Gobet}}, \bibinfo
  {author} {\bibfnamefont {D.}~\bibnamefont {Denis-Petit}}, \bibinfo {author}
  {\bibfnamefont {F.}~\bibnamefont {Hannachi}}, \bibinfo {author}
  {\bibfnamefont {V.}~\bibnamefont {Meot}}, \ and\ \bibinfo {author}
  {\bibfnamefont {M.}~\bibnamefont {Tarisien}},\ }\href@noop {} {\bibfield
  {journal} {\bibinfo  {journal} {Journal of Applied Physics}\ }\textbf
  {\bibinfo {volume} {119}},\ \bibinfo {pages} {013301} (\bibinfo {year}
  {2016})}\BibitemShut {NoStop}%
\bibitem [{\citenamefont {Gitomer}\ \emph {et~al.}(1986)\citenamefont
  {Gitomer}, \citenamefont {Jones}, \citenamefont {Begay}, \citenamefont
  {Ehler}, \citenamefont {Kephart},\ and\ \citenamefont
  {Kristal}}]{gitomer1986fast}%
  \BibitemOpen
  \bibfield  {author} {\bibinfo {author} {\bibfnamefont {S.}~\bibnamefont
  {Gitomer}}, \bibinfo {author} {\bibfnamefont {R.}~\bibnamefont {Jones}},
  \bibinfo {author} {\bibfnamefont {F.}~\bibnamefont {Begay}}, \bibinfo
  {author} {\bibfnamefont {A.}~\bibnamefont {Ehler}}, \bibinfo {author}
  {\bibfnamefont {J.}~\bibnamefont {Kephart}}, \ and\ \bibinfo {author}
  {\bibfnamefont {R.}~\bibnamefont {Kristal}},\ }\href@noop {} {\bibfield
  {journal} {\bibinfo  {journal} {The Physics of fluids}\ }\textbf {\bibinfo
  {volume} {29}},\ \bibinfo {pages} {2679} (\bibinfo {year}
  {1986})}\BibitemShut {NoStop}%
\bibitem [{\citenamefont {Humphries}(2013)}]{humphries2013charged}%
  \BibitemOpen
  \bibfield  {author} {\bibinfo {author} {\bibfnamefont {S.}~\bibnamefont
  {Humphries}},\ }\href@noop {} {\emph {\bibinfo {title} {Charged particle
  beams}}}\ (\bibinfo  {publisher} {Courier Corporation},\ \bibinfo {year}
  {2013})\BibitemShut {NoStop}%
\bibitem [{\citenamefont {Verboncoeur}\ \emph {et~al.}(1995)\citenamefont
  {Verboncoeur}, \citenamefont {Langdon},\ and\ \citenamefont
  {Gladd}}]{verboncoeur1995object}%
  \BibitemOpen
  \bibfield  {author} {\bibinfo {author} {\bibfnamefont {J.~P.}\ \bibnamefont
  {Verboncoeur}}, \bibinfo {author} {\bibfnamefont {A.~B.}\ \bibnamefont
  {Langdon}}, \ and\ \bibinfo {author} {\bibfnamefont {N.}~\bibnamefont
  {Gladd}},\ }\href@noop {} {\bibfield  {journal} {\bibinfo  {journal}
  {Computer Physics Communications}\ }\textbf {\bibinfo {volume} {87}},\
  \bibinfo {pages} {199} (\bibinfo {year} {1995})}\BibitemShut {NoStop}%
\bibitem [{\citenamefont {Child}(1911)}]{child1911discharge}%
  \BibitemOpen
  \bibfield  {author} {\bibinfo {author} {\bibfnamefont {C.}~\bibnamefont
  {Child}},\ }\href@noop {} {\bibfield  {journal} {\bibinfo  {journal}
  {Physical Review (Series I)}\ }\textbf {\bibinfo {volume} {32}},\ \bibinfo
  {pages} {492} (\bibinfo {year} {1911})}\BibitemShut {NoStop}%
\bibitem [{\citenamefont {Langmuir}(1923)}]{langmuir1923effect}%
  \BibitemOpen
  \bibfield  {author} {\bibinfo {author} {\bibfnamefont {I.}~\bibnamefont
  {Langmuir}},\ }\href@noop {} {\bibfield  {journal} {\bibinfo  {journal}
  {Physical Review}\ }\textbf {\bibinfo {volume} {21}},\ \bibinfo {pages} {419}
  (\bibinfo {year} {1923})}\BibitemShut {NoStop}%
\bibitem [{\citenamefont {Agostinelli}\ \emph {et~al.}(2003)\citenamefont
  {Agostinelli}, \citenamefont {Allison}, \citenamefont {Amako}, \citenamefont
  {Apostolakis}, \citenamefont {Araujo}, \citenamefont {Arce}, \citenamefont
  {Asai}, \citenamefont {Axen}, \citenamefont {Banerjee}, \citenamefont
  {Barrand} \emph {et~al.}}]{agostinelli2003geant4}%
  \BibitemOpen
  \bibfield  {author} {\bibinfo {author} {\bibfnamefont {S.}~\bibnamefont
  {Agostinelli}}, \bibinfo {author} {\bibfnamefont {J.}~\bibnamefont
  {Allison}}, \bibinfo {author} {\bibfnamefont {K.~a.}\ \bibnamefont {Amako}},
  \bibinfo {author} {\bibfnamefont {J.}~\bibnamefont {Apostolakis}}, \bibinfo
  {author} {\bibfnamefont {H.}~\bibnamefont {Araujo}}, \bibinfo {author}
  {\bibfnamefont {P.}~\bibnamefont {Arce}}, \bibinfo {author} {\bibfnamefont
  {M.}~\bibnamefont {Asai}}, \bibinfo {author} {\bibfnamefont {D.}~\bibnamefont
  {Axen}}, \bibinfo {author} {\bibfnamefont {S.}~\bibnamefont {Banerjee}},
  \bibinfo {author} {\bibfnamefont {G.}~\bibnamefont {Barrand}},  \emph
  {et~al.},\ }\href@noop {} {\bibfield  {journal} {\bibinfo  {journal} {Nuclear
  instruments and methods in physics research section A: Accelerators,
  Spectrometers, Detectors and Associated Equipment}\ }\textbf {\bibinfo
  {volume} {506}},\ \bibinfo {pages} {250} (\bibinfo {year}
  {2003})}\BibitemShut {NoStop}%
\bibitem [{\citenamefont {Allison}\ \emph {et~al.}(2016)\citenamefont
  {Allison}, \citenamefont {Amako}, \citenamefont {Apostolakis}, \citenamefont
  {Arce}, \citenamefont {Asai}, \citenamefont {Aso}, \citenamefont {Bagli},
  \citenamefont {Bagulya}, \citenamefont {Banerjee}, \citenamefont {Barrand}
  \emph {et~al.}}]{allison2016recent}%
  \BibitemOpen
  \bibfield  {author} {\bibinfo {author} {\bibfnamefont {J.}~\bibnamefont
  {Allison}}, \bibinfo {author} {\bibfnamefont {K.}~\bibnamefont {Amako}},
  \bibinfo {author} {\bibfnamefont {J.}~\bibnamefont {Apostolakis}}, \bibinfo
  {author} {\bibfnamefont {P.}~\bibnamefont {Arce}}, \bibinfo {author}
  {\bibfnamefont {M.}~\bibnamefont {Asai}}, \bibinfo {author} {\bibfnamefont
  {T.}~\bibnamefont {Aso}}, \bibinfo {author} {\bibfnamefont {E.}~\bibnamefont
  {Bagli}}, \bibinfo {author} {\bibfnamefont {A.}~\bibnamefont {Bagulya}},
  \bibinfo {author} {\bibfnamefont {S.}~\bibnamefont {Banerjee}}, \bibinfo
  {author} {\bibfnamefont {G.}~\bibnamefont {Barrand}},  \emph {et~al.},\
  }\href@noop {} {\bibfield  {journal} {\bibinfo  {journal} {Nuclear
  Instruments and Methods in Physics Research Section A: Accelerators,
  Spectrometers, Detectors and Associated Equipment}\ }\textbf {\bibinfo
  {volume} {835}},\ \bibinfo {pages} {186} (\bibinfo {year}
  {2016})}\BibitemShut {NoStop}%
\bibitem [{\citenamefont {Souici}\ \emph {et~al.}(2017)\citenamefont {Souici},
  \citenamefont {Khalil}, \citenamefont {Muller}, \citenamefont {Raffy},
  \citenamefont {Barillon}, \citenamefont {Belafrites}, \citenamefont
  {Champion},\ and\ \citenamefont {Fromm}}]{souici2017single}%
  \BibitemOpen
  \bibfield  {author} {\bibinfo {author} {\bibfnamefont {M.}~\bibnamefont
  {Souici}}, \bibinfo {author} {\bibfnamefont {T.~T.}\ \bibnamefont {Khalil}},
  \bibinfo {author} {\bibfnamefont {D.}~\bibnamefont {Muller}}, \bibinfo
  {author} {\bibfnamefont {Q.}~\bibnamefont {Raffy}}, \bibinfo {author}
  {\bibfnamefont {R.}~\bibnamefont {Barillon}}, \bibinfo {author}
  {\bibfnamefont {A.}~\bibnamefont {Belafrites}}, \bibinfo {author}
  {\bibfnamefont {C.}~\bibnamefont {Champion}}, \ and\ \bibinfo {author}
  {\bibfnamefont {M.}~\bibnamefont {Fromm}},\ }\href@noop {} {\bibfield
  {journal} {\bibinfo  {journal} {The Journal of Physical Chemistry B}\
  }\textbf {\bibinfo {volume} {121}},\ \bibinfo {pages} {497} (\bibinfo {year}
  {2017})}\BibitemShut {NoStop}%
\bibitem [{\citenamefont {Hahn}\ \emph {et~al.}(2017)\citenamefont {Hahn},
  \citenamefont {Meyer}, \citenamefont {Kunte}, \citenamefont {Solomun},\ and\
  \citenamefont {Sturm}}]{hahn2017measurements}%
  \BibitemOpen
  \bibfield  {author} {\bibinfo {author} {\bibfnamefont {M.~B.}\ \bibnamefont
  {Hahn}}, \bibinfo {author} {\bibfnamefont {S.}~\bibnamefont {Meyer}},
  \bibinfo {author} {\bibfnamefont {H.-J.}\ \bibnamefont {Kunte}}, \bibinfo
  {author} {\bibfnamefont {T.}~\bibnamefont {Solomun}}, \ and\ \bibinfo
  {author} {\bibfnamefont {H.}~\bibnamefont {Sturm}},\ }\href@noop {}
  {\bibfield  {journal} {\bibinfo  {journal} {Physical Review E}\ }\textbf
  {\bibinfo {volume} {95}},\ \bibinfo {pages} {052419} (\bibinfo {year}
  {2017})}\BibitemShut {NoStop}%
\end{thebibliography}%

\end{document}